\documentclass[11pt,a4paper]{article}
\pdfoutput=1

\usepackage{jheppub}
\usepackage{booktabs}			
\usepackage{multirow}			
\usepackage{tabularx,blindtext,booktabs,ragged2e} 

\def\eps{\epsilon}

\def\e{\epsilon}
\def\d{\hbox{d}}

\def\permille{\ensuremath{{}^\text{o}\mkern-5mu/\mkern-3mu_\text{oo}}}

\preprint{
ZU--TH 10/15}

\title{The rare decay $H\to Z\gamma$ in perturbative QCD}
    
\author{T.\ Gehrmann, S.\ Guns  and D.\ Kara}

\affiliation{
Physik-Institut, Universit\"at Z\"urich, Winterthurerstrasse 190, 8057 Z\"urich, 
Switzerland}

\abstract{The rare Higgs boson decay $H\to Z\gamma$ is forbidden at tree-level. In the 
Standard Model, it is loop-mediated through a $W$ boson or a heavy quark. We analytically 
compute the QCD correction to the heavy quark loop, confirming earlier purely numerical results,
that were obtained for on-shell renormalization.  The small quark mass expansion of the decay 
matrix element contains only single-logarithmic contributions at each perturbative order, which is in contrast to the double 
logarithms observed in $H\to \gamma\gamma$.  We
investigate the numerical interplay of bottom and top quark contributions, and the dependence of the 
result on the renormalization scheme.}

\begin{document}
\maketitle

\section{Introduction}
The discovery of the Higgs boson~\cite{Aad:2012tfa,Chatrchyan:2012ufa} 
initiated a large research program aiming to determine the 
Higgs boson properties and to establish the details of the Higgs mechanism  of electroweak symmetry breaking. 
Experimental measurements  are up to now consistent with expectations from the Standard Model of 
particle physics. It is anticipated that the upcoming LHC data taking period at higher energy and luminosity 
will provide more precise measurements and open up new observables that were previously inaccessible. 
In this context, the different decay modes of the Higgs boson play a very vital role. The decays to massive gauge bosons 
and to fermions are allowed at tree-level and provide direct measurements of the Higgs boson couplings.  The rare 
decays $H\to \gamma\gamma$ and $H\to Z\gamma$ are forbidden at tree-level, they are mediated through 
loops containing massive particles~\cite{Ellis:1975ap,Cahn:1978nz,Bergstrom:1985hp}. 
As such, they are more sensitive to new physics effects from high energy scales 
than the tree-level dominated decay modes.

The $H\to \gamma\gamma$ decay mode was among the most significant signatures in the Higgs boson 
discovery~\cite{Aad:2012tfa,Chatrchyan:2012ufa}, it has been measured 
in the meantime~\cite{Aad:2014eha,Khachatryan:2014ira} with a relative precision of below twenty per cent. 
The branching ratio for $H\to Z\gamma$, including the leptonic branching ratio of the $Z$ boson, is considerably lower and only upper bounds could be established on it 
up to now~\cite{Aad:2014fia,Chatrchyan:2013vaa}. Once established, this decay will provide access to a broader 
spectrum of observables than $H\to \gamma \gamma$, since the decay of the $Z$ boson to leptons will enable the 
study of spin-dependent particle correlations. It should be noted that the decay $H\to Z\gamma$ will be 
identified through a 
mass-cut on the pair of decay leptons, and that it should be considered to be a pseudo-observable~\cite{Passarino:2013nka}. 

Higher order QCD corrections to $H\to Z\gamma$ from gluon exchange in the 
top quark loop were derived in~\cite{Spira:1991tj} by performing a purely numerical evaluation of the 
relevant two-loop integrals in terms of five-dimensional Feynman parameter representations. 
The results derived in~\cite{Spira:1991tj} use an on-shell renormalization for the top quark mass and the Yukawa coupling. 
Electroweak 
corrections to this decay are not known at present~\cite{Denner:2011mq,Passarino:2013nka}. It is the aim of this paper to 
rederive the QCD corrections to  $H\to Z\gamma$ in an analytical form and to quantify uncertainties on them 
arising from scheme and scale dependence. 

Besides its phenomenological implications for  $H\to Z\gamma$, our calculation also provides an important subset of 
two-loop integrals relevant to the two-loop amplitudes $gg\to Hg$ and $qg \to Hq$ with full top quark mass dependence. 
These amplitudes are known at present only at one loop~\cite{Spira:1995rr}, corresponding to the leading order 
in perturbation theory. For precision studies of the transverse momentum distribution of the Higgs boson 
and of Higgs-boson-plus-jet production, an effective field theory in the limit of infinite top quark mass is 
used commonly. In this approach, NLO QCD corrections were derived~\cite{Schmidt:1997wr,Glosser:2002gm,deFlorian:1999zd,Ravindran:2002dc}
and the calculation of 
the NNLO  corrections is well-advanced~\cite{Gehrmann:2011aa,Boughezal:2013uia,Chen:2014gva,Boughezal:2015dra,Boughezal:2015aha}. 
The effective field theory description is however inappropriate at 
large transverse momenta, where the top quark loop is resolved by the recoiling jet; and it is precisely in this 
region that deviations from the Standard Model due to new heavy states could become visible. The calculation of NLO 
QCD corrections with exact top quark mass dependence is therefore recognized as high-priority 
aim~\cite{Dittmaier:2012vm,Heinemeyer:2013tqa}, and the integrals derived here will be an important step towards it. 

This paper is structured as follows: in Section~\ref{sec:notation}, we establish the notation and discuss the different 
contributions to the $H\to Z\gamma$ decay. Section~\ref{sec:calc} describes the calculation of the 
amplitude, including a detailed discussion of the relevant two-loop three-point integrals and of the renormalization. 
The numerical results are discussed in Section~\ref{sec:results}, and we conclude with Section~\ref{sec:conc}.

\section{The $H\to Z\gamma$ decay in the Standard Model}
\label{sec:notation}

The Standard Model does not allow a tree-level coupling of the Higgs boson to a photon and a $Z$ boson. The process
\begin{displaymath}
H(q) \to Z(p_1) \gamma(p_2)
\end{displaymath}
is mediated through a virtual particle loop, containing either a $W$ boson or a 
massive quark~\cite{Cahn:1978nz,Bergstrom:1985hp}. 
The Lorentz 
structure of its Feynman amplitude 
is constrained by gauge invariance to contain only a single scalar form factor:
\begin{equation}
{\cal M} = A\, \epsilon_{1,\mu}(p_1,\lambda_1) \, \epsilon_{2,\nu}(p_2,\lambda_2) \, \frac{P^{\mu\nu}}{P^2} \,,
\label{eq:tensamp}
\end{equation}
where the projector $P^{\mu\nu}$ is given by
\begin{equation}
P^{\mu\nu} = p_2^\mu p_1^\nu - (p_1\cdot p_2) g^{\mu \nu} \,.
\end{equation}
The decay width $H\to Z\gamma$ is obtained as
\begin{equation}
\Gamma = \frac{G_F^2 \, \alpha \, m_W^2}{4 \, m_H^3 \left(m_H^2-m_Z^2\right)} \, |A|^2
\label{eq:gamma}
\end{equation}
with Fermi's coupling constant $G_F$, the fine-structure constant $\alpha$ and the Higgs and $Z$~boson masses $m_H$ and $m_Z$, respectively. Depending on the particle coupled to the external Higgs boson, the amplitude can be further decomposed into contributions from the $W$ boson and the fermions $q$:
\begin{equation}
A = c_W A_W + \sum_q c_q A_q\,.
\end{equation}
The coupling factors are
\begin{equation}
c_W = \mathrm{cos} \, \theta_w \,, \qquad c_q = N_c \, \frac{2 \, Q_q \left(I^3_q - 2 \, Q_q \, \mathrm{sin}^2 \, \theta_w \right)}{\mathrm{cos} \, \theta_w}  \;,
\end{equation}
where $\theta_w$ is the weak mixing angle, $N_c$ the number of colors, $Q_q$ the charge of the fermion and $I^3_q$ the third component of its weak isospin. 
Due to the mass hierarchy of the particles involved, we will only consider the dominant pieces coming from the $W$ boson, the top quark and the bottom quark.

The Born-level contribution to the amplitude arises at one loop, it is written as:
\begin{equation} 
A^{(1)} = c_W A^{(1)}_W + c_t A^{(1)}_t + c_b A^{(1)}_b\,.
\end{equation}
Higher-order perturbative corrections are obtained by a loop expansion of the amplitude $A$. Next-to-leading order 
QCD corrections affect only $A_t$ and $A_b$, they correspond to two-loop graphs with an internal mass:
\begin{equation}
A_q(m_H,m_Z,m_q,\alpha_s,\mu) = A^{(1)}_q (m_H,m_Z,m_q) + \frac{\alpha_s(\mu)}{\pi} A^{(2)}_q(m_H,m_Z,m_q,\mu) \,.
\label{eq:amp}
\end{equation}
For the dominant top quark contribution, the two-loop correction $A^{(2)}_t$
 was computed numerically (based on the Feynman parameter representation of the 
 amplitude) using an on-shell renormalization for the 
quark mass and the Yukawa coupling in~\cite{Spira:1991tj}. 

\section{Calculation of the amplitude}
\label{sec:calc}
To compute the one-loop amplitudes $A^{(1)}_W$ and $A^{(1)}_q$ as well as the two-loop QCD contribution $A^{(2)}_q$ to the quark-mediated amplitude for $H\to Z\gamma$, we 
project all relevant Feynman diagrams generated by \textsc{Qgraf}~\cite{qgraf} onto the tensor structure~(\ref{eq:tensamp}) using \textsc{Form}~\cite{Kuipers:2012rf}. The resulting Feynman integrals are reduced to a set of master integrals 
with the help of integration-by-parts (IBP) identities~\cite{Chetyrkin:1981qh}, which are solved using the Laporta 
algorithm~\cite{Laporta:2001dd} implemented in the \textsc{Reduze2} code~\cite{vonManteuffel:2012np,fermat}. After the 
reduction, the amplitude can be expressed in terms of a certain number of master integrals depending on the loop order.

Each of these master integrals has a specific mass dimension, which can be scaled out by multiplying with the 
appropriate power of the mass $m_i$ running in the loop, such that the resulting dimensionless integrals are only functions of the mass ratios
$m_H^2/m_i^2$ and $m_Z^2/m_i^2$. We parametrize this dependence by introducing Landau-type variables
\begin{equation}
m_H^2 = - m_i^2 \frac{(1-h_i)^2}{h_i}\,, \qquad m_Z^2 = -m_i^2 \frac{(1-z_i)^2}{z_i} \qquad \left(i=W,q\right)\,.
\label{eq:landau}
\end{equation}

For the sake of readability, we drop the subscript $q$ of these variables in the following whenever we deal with quark-mediated amplitudes, i.e. $h_q \equiv h$, $z_q \equiv z$.

\subsection{The one-loop amplitude}
\begin{figure}[t]
\begin{center}
\includegraphics[width=0.92\textwidth]{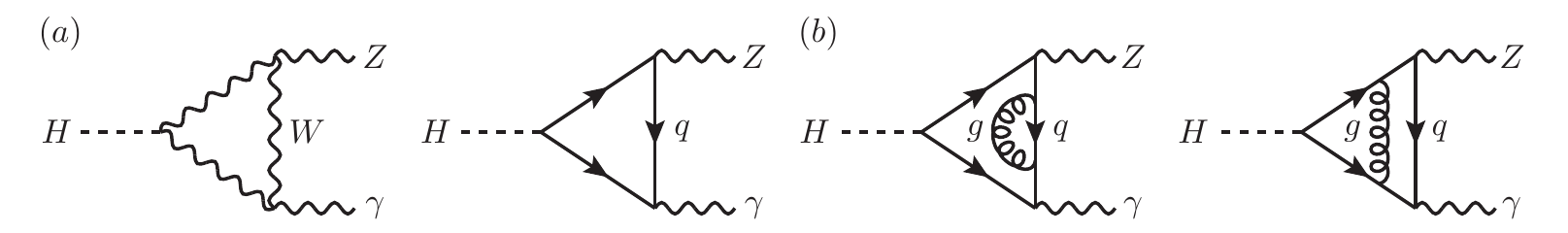}
\caption{Example diagrams $(a)$ for the computation of the one-loop amplitudes $A^{(1)}_W$ and $A^{(1)}_q$ and $(b)$ for the computation of the two-loop amplitude $A^{(2)}_q$.}
\label{fig:diagrams}
\end{center}
\end{figure}

The Born-level one-loop amplitudes for the contributions of $W$-bosons $A^{(1)}_W$ and 
heavy quarks $A^{(1)}_q$
to  $H\to Z\gamma$ were derived in~\cite{Cahn:1978nz,Bergstrom:1985hp}. Example diagrams are depicted in  
Fig.~\ref{fig:diagrams}$(a)$.

In terms of the Landau variables introduced above, these results read
\begin{align}
A^{(1)}_W = \frac{4\,i\,S_\e\,m_W^4}{h_Wz_W^2} &\left[\left\{(h_W^2+1) (z_W^2+1) - 4 h_W (z_W+1)^2 \right\} \, \left\{ \frac{(h_W-z_W) (h_Wz_W-1)}{h_W} \right. \right. \nonumber \\
	&\left. \quad- \frac{(h_W+1) (z_W-1)^2}{h_W-1} \, \log(h_W) + (z_W^2-1) \, \log(z_W) \right\} \nonumber \\
	&\,+\left\{ (z_W + z_W^2 (z_W+4)) (h_W^2+1) - 2h_W (z_W^2-1)^2 \right\} \, \nonumber \\
	&\,\left. \quad \left\{ \log^2(h_W) - \log^2(z_W) \right\} \right] + {\cal O}(\eps) \,, \\
A^{(1)}_q = i\,S_\e\,m_q^3\,y_q\,v&\left[8\,\left\{h+\frac{1}{h}-\left(z+\frac{1}{z}\right)\right.\right. \nonumber \\
	&\left. \qquad- \frac{(h+1) (z-1)^2}{z(h-1)} \, \log(h) + \left(z-\frac{1}{z}\right) \, \log(z) \right\} \nonumber \\
	&\,\left.-2\left\{h+\frac{1}{h}-z-\frac{1}{z}+4\right\} \left\{ \log^2(h) - \log^2(z) \right\} \right] + {\cal O}(\eps)
\label{eq:LO}
\end{align}
with the Standard Model Higgs vacuum expectation value $v$ and the Yukawa coupling
\begin{equation}
y_q = \frac{m_q}{v}
\end{equation}
associated with the quark $q$. The normalization factor
\begin{equation}
S_\e = \frac{i \, \Gamma\left(1+\e\right)}{16\pi^2} \, \left(\frac{4\pi \mu_0^2}{m_i^2}\right)^\e
\label{eq:norm}
\end{equation}
arises from the integration measure $\int{\rm d}^D k/(2\pi)^D$ of the master integrals in $D=4-2\e$ dimensions, where $\mu_0$ is the mass scale of dimensional regularization.

\subsection{Differential equations and integral basis}
The two-loop amplitude for the quark contribution $A^{(2)}_q$ has been computed purely numerically in terms of 
a five-dimensional Feynman parameter integral in~\cite{Spira:1991tj}. We derive an analytical expression for this 
amplitude, through a reduction of all two-loop integrals to a set of master integrals. 

To compute the two-loop master integrals, we use the method of differential equations~\cite{Kotikov:1990kg,Remiddi:1997ny,Gehrmann:1999as,Henn:2013pwa}. In this method, differential equations in 
internal masses and external invariants are derived for each integral by performing the differentiation on the 
integrand, which is then related to the original master integral by the IBP identities. With this, we obtain inhomogeneous 
differential equations in either Landau variable, plus a trivial homogeneous equation in $m_q$ for each integral. 
The differential equations are solved in a bottom-up approach, i.e. starting from the master integrals with the lowest number of different propagators (`topology') because they will show up in differential equations of higher topologies.

The coefficients of the individual master integrals in the homogeneous and inhomogeneous terms of the differential 
equations are rational functions of $h$ and $z$. Upon partial fractioning, only a limited number of polynomials in
$h$ and $z$ appear. These form the so-called alphabet associated with this set of master integrals:
\begin{equation}
\{ l_1,\ldots, l_{12} \} 
\label{eq:denom}
\end{equation}
with
\begin{align*}
l_1 &= z \,, \\
l_2 &= z+1 \,, \\
l_3 &= z-1 \,, \\
l_4 &= h \,, \\
l_5 &= h+1 \,, \\
l_6 &= h-1 \,, \\
l_7 &= h-z \,, \\
l_8 &= h z-1 \,, \\
l_9 &= h^2-h z-h+1 \,, \\
l_{10} &= h^2 z-h z-h+z \,, \\
l_{11} &= z^2-h z-z+1 \,, \\
l_{12} &= z^2 h-h z-z+h \,.
\end{align*}

\begin{figure}[!ht]
\begin{center}
\includegraphics[width=0.92\textwidth]{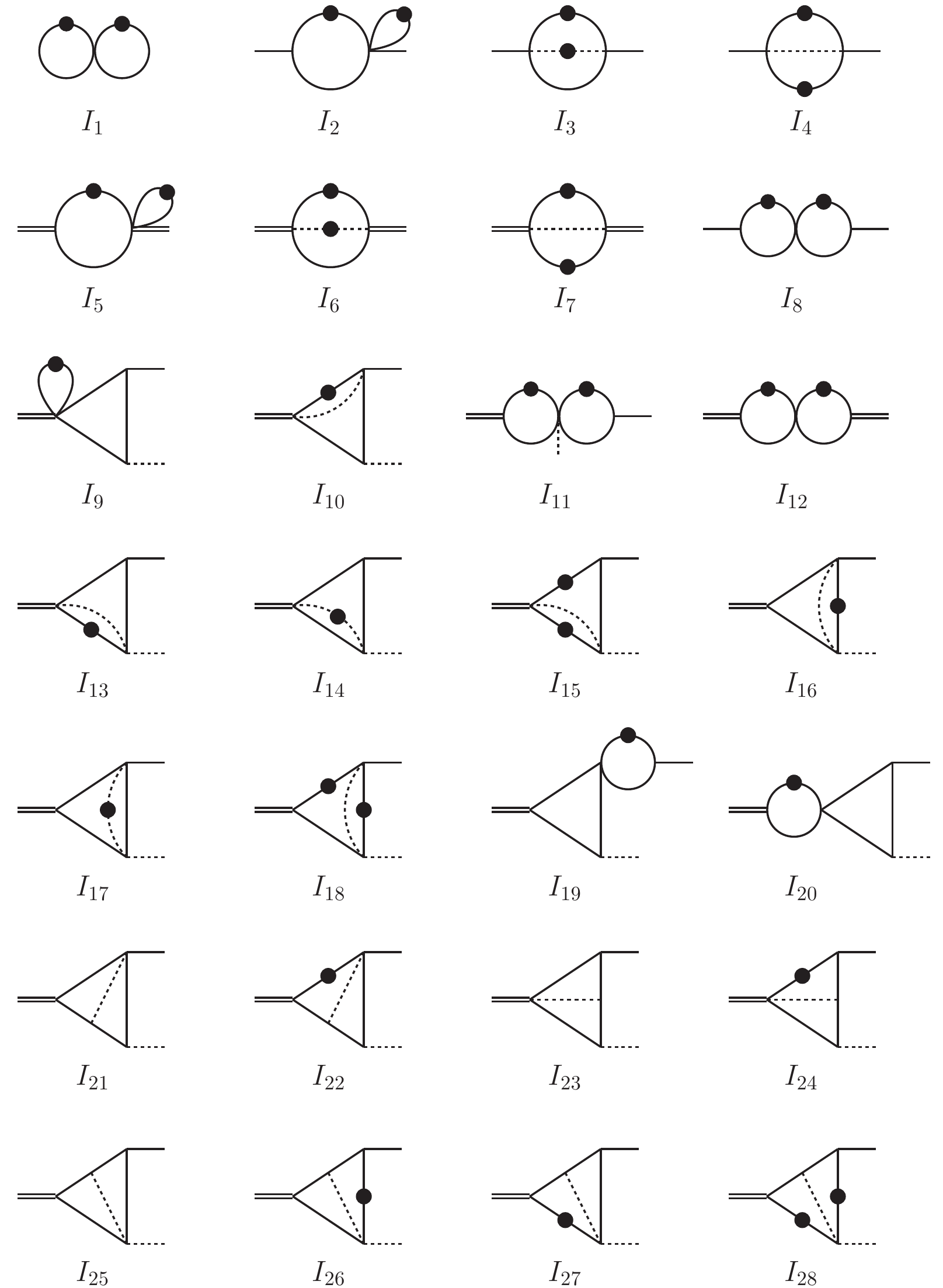}
\caption{Two-loop master integrals for the calculation of $A^{(2)}_q$. Dashed lines are massless, whereas internal solid lines denote propagators with mass $m_q$. Double and solid external lines correspond to virtualities $m_H^2$ and $m_Z^2$, respectively. Dotted propagators are taken to be squared.}
\label{fig:master}
\end{center}
\end{figure}

With an appropriate choice of basis integrals~\cite{Henn:2013pwa}, the full system of differential equations for  
all 28 master integrals (written as 28-component vector $\vec{M}$) takes the form of a total 
differential,
\begin{equation}
\d \vec{M}(h,z) = \e \sum_{k=1}^{12} R_k \, \d \log (l_k) \, \vec{M} (h,z)\,,
\label{eq:canon}
\end{equation}
where the matrices $R_k$ contain only rational numbers. In this case, two important features can be exploited.  First, the differential equations can be integrated order by order in $\e$ in terms of generalized harmonic polylogarithms (GHPLs)~\cite{Goncharov:1998kja,Remiddi:1999ew,Gehrmann:2000zt,Vollinga:2004sn}. They are defined as iterated integrals according to
\begin{align}
G\left(w_1,\ldots,w_n;x\right) &\equiv \int_0^x \d t \, \frac{1}{t-w_1} \, G\left(w_2,\ldots,w_n;t\right) \,, \nonumber \\
G\left(\vec{0}_n;x\right) &\equiv \frac{\log^nx}{n!} \,,
\end{align}
with indices $w_i$ and argument $x$.
Second, the results will be expressed as a linear combination of GHPLs of homogeneous weight. We arrive at a total differential of the form (\ref{eq:canon}) starting from the Laporta basis $\vec{I}$ depicted in Fig.~\ref{fig:master} and subsequently applying the algorithm described in~\cite{Gehrmann:2014bfa}. The resulting canonical basis $\vec{M}$ in terms of the Laporta basis $\vec{I}$ is given in Appendix~\ref{sec:appendix}.

Since the alphabet~(\ref{eq:denom}) is not linear in the Landau variables, we further decompose it to enable the integration in either $h$ or $z$, which yields a solution up to an integration constant that only depends on the other variable. This boundary value is then determined by imposing regularity in special kinematic points, where the integrals are known to be regular from physical arguments. In our case, these points are given by
\begin{align}
h = 1 \qquad &\leftrightarrow \qquad m_H^2 = 0 \,, \nonumber \\
z = 1 \qquad &\leftrightarrow \qquad m_Z^2 = 0 \,, \nonumber \\
h = z \qquad &\leftrightarrow \qquad m_H^2 = m_Z^2 \,, \nonumber \\
h = \frac{1}{z} \qquad &\leftrightarrow \qquad m_H^2 = m_Z^2 \,,
\end{align}
i.e. they correspond to the limit where the masses of the external particles either vanish or coincide. Choosing one of these points such that the rational prefactors in Eqs.~(\ref{eq:appendix}) are equal to zero considerably reduces the complexity of the integration constants.

By taking limits in these kinematic points, we are left with GHPLs that contain the same variable $x\in\{h,z\}$ both in the argument and in the indices. In order to simplify the result and to obtain a unique representation, we use an inhouse \textsc{Mathematica} implementation~\cite{Gehrmann:2013cxs,weihs}
 relying on the symbol and coproduct formalism~\cite{Duhr:2012fh} to transform GHPLs of the type
\begin{equation}
G\left(w_1(x),\ldots,w_n(x);x\right) \to G\left(a_1,\ldots,a_n;x\right) \,,
\label{eq:trafo1}
\end{equation}
where the transformed indices $a_i$ are nothing but complex numbers. In doing so, we end up with GHPLs up to weight four, which are given by
\begin{align}
G\left(a_1,\ldots,a_n;h\right) \quad &\text{with} \quad a_i \in \{ 0,\pm1,z,\frac{1}{z},J_z,\frac{1}{J_z},K_z^\pm, L_z^\pm \} \,, \nonumber \\
G\left(b_1,\ldots,b_n;z\right) \quad &\text{with} \quad b_i \in \{ 0,\pm1,c,\bar{c} \} \,,
\label{eq:alphabet}
\end{align}
where 
\begin{align}
c &= \frac{1}{2} \left(1 + i\sqrt{3}\right) \,, \nonumber \\
J_z &= \frac{z}{1 - z + z^2} \,, \nonumber \\
K_z^\pm &= \frac{1}{2} \left(1 + z \pm \sqrt{\text{--}3 + 2\,z + z^2}\right) \,, \nonumber \\
L_z^\pm &= \frac{1}{2\,z} \left(1 + z \pm \sqrt{1 + 2\,z - 3\,z^2}\right)
\end{align}
for the underlying set of master integrals. Further transformations of the type
\begin{equation}
G\left(w_1(x),\ldots,w_n(x);y\right) \to G\left(c_1(y),\ldots,c_n(y);x\right)
\label{eq:trafo2}
\end{equation}
become necessary when the integration is performed in a different variable compared to the integration of a master integral of a subtopology which enters the differential equation under consideration.

It remains to comment on one issue: Transformations of the types (\ref{eq:trafo1}) and (\ref{eq:trafo2}) are performed for all master integrals except for $M_{25}\text{--}M_{28}$, where GHPLs of the form
\begin{equation}
G\left(w_1(x),\ldots,w_n(x);x\right)
\label{eq:nontrans}
\end{equation}
with $w_i=\{K_x^\pm,L_x^\pm\}$ occur. The transformations were not necessary in this case because the results of these integrals do not enter the differential equation of any master integral of higher topology. However, transformations of these types could become desirable when four-point functions are computed, where these four coupled master integrals appear as subtopologies. This could be attempted by a multivariate extension of the generalized weights approach described in~\cite{vonManteuffel:2013vja}, i.e. rather working on non-linear indices of the form~(\ref{eq:denom}) than on linear ones as in Eq.~(\ref{eq:alphabet}).

We would like to state that the results of the master integrals were checked in several ways. We performed transformations of the type (\ref{eq:trafo2}) and verified that the solution fulfills the differential equation in the other variable. This check works only up to a constant, which is why we compared each master integral numerically against \textsc{SecDec}~\cite{Borowka:2015mxa} and found agreement to high precision. The analytic expressions of the master integrals are rather lengthy and will not be reproduced here. They are available in \textsc{Mathematica} and \textsc{Form} format together with the arXiv submission of this paper.

\subsection{Calculation of the two-loop amplitude}
For the two-loop amplitude, six generic Feynman diagrams and their permutations have to be evaluated. They emerge from the one-loop diagram in Fig.~\ref{fig:diagrams}$(a)$ by attaching one gluon propagator to the fermion lines in every possible way, which is depicted in Fig.~\ref{fig:diagrams}$(b)$. After the manipulations described in the beginning of Section~\ref{sec:calc} and after inserting the analytic results of the master integrals, we are left with the unrenormalized two-loop amplitude.

For its renormalization, we consider three different prescriptions:
\begin{itemize}
\item[(a)] quark mass and Yukawa coupling in the on-shell (OS) scheme.
\item[(b)] quark mass in the OS scheme, Yukawa coupling in the $\overline{{\rm MS}}$ scheme.
\item[(c)] quark mass and Yukawa coupling in the $\overline{{\rm MS}}$ scheme.
\end{itemize}
Since the two-loop amplitude is the leading order in $\alpha_s$, no renormalization of the 
gauge coupling is required. 

All three prescriptions yield the same pole parts of the renormalization counter terms and produce finite expressions for the renormalized amplitude. They are related by finite scheme transformations, which is why we choose to compute the renormalized amplitude in scheme (a) and use it to derive the results in schemes (b) and (c). In the pure OS scheme, the quantity
\begin{equation}
\frac{1}{m_q}\,\delta m_{\mathrm{OS}}\,C^{(1)}_q + Z_{\mathrm{OS}}\,A^{(1)}_q
\label{eq:renormOS}
\end{equation}
has to be added to the unrenormalized two-loop amplitude in order to remove its divergences, where $Z_{\mathrm{OS}}$ and $\delta m_{\mathrm{OS}}$ are the Yukawa and mass renormalization constants, respectively \cite{Bernreuther:2004ih}:
\begin{align}
Z_{\mathrm{OS}} &= \frac{\alpha_s(\mu)}{\pi} 16\,i\,\pi^2 \,S_\e \, \frac{C_F}{4} \, \frac{3-2\e}{\e\left(1-2\e\right)}\,, \\
\delta m_{\mathrm{OS}} &= m_q \, Z_{\mathrm{OS}} \, .
\end{align}
Note that Eq.~(\ref{eq:renormOS}) requires the calculation of the one-loop amplitude $A^{(1)}_q$ and the mass counterterm $C^{(1)}_q$ up to $\mathcal{O}(\e)$. $C^{(1)}_q$ can be computed from the sum of the three diagrams that are obtained by 
putting a mass insertion into one of the fermion lines in Fig.~\ref{fig:diagrams}$(a)$.

Next, we express the OS quantities $M_q$ and $Y_q$ in terms of $\overline{{\rm MS}}$ quantities 
$\overline{m}_q$ and $\overline{y}_q$ at a particular matching scale $\mu_m$ using the standard relations (e.g.~\cite{Melnikov:2000qh,Chetyrkin:1999qi,Jamin:1997rt}):
\begin{align}
M_q &= \overline{m}_q(\mu) \, \left( 1 + \Delta \right) \,, \nonumber \\
Y_q &= \overline{y}_q(\mu) \, \left( 1 + \Delta \right) \,, \nonumber \\
\Delta &= \frac{\alpha_s(\mu)}{\pi} \, C_F \left(1 + \frac{3}{4} \, \log \frac{\mu^2}{\overline{m}_q^2(\mu)} \right) \,.
\end{align}
We perform this matching at the scale of the running $\overline{{\rm MS}}$ quark mass $\overline{m}_q$. Starting from the OS result $A^{(2,a)}_q(m_H,m_Z,M_q)$, these scheme transformations 
induce finite shifts in the amplitudes of the prescriptions (b) and (c):
\begin{eqnarray} 
A^{(2,b)}_q(m_H,m_Z,\overline{m}_q,\mu) &=& A^{(2,a)}_q(m_H,m_Z,\overline{m}_q(\mu)) +  \Delta \cdot \, A^{(1)}_q(m_H,m_Z,\overline{m}_q(\mu)) \nonumber \,,\\
A^{(2,c)}_q(m_H,m_Z,\overline{m}_q,\mu) &=& A^{(2,b)}_q(m_H,m_Z,\overline{m}_q(\mu)) +  \overline{\Delta} \cdot
\left. \frac{\partial  A^{(1)}_q(m_H,m_Z,M_q)}{\partial
M_q} \right|_{M_q=\overline{m}_q(\mu)} \,.
\end{eqnarray}
In practice, the coefficient $\overline{\Delta}$ emerges by making the following replacements in Eq.~(\ref{eq:LO}), where the Landau variables $\bar{h}$ and $\bar{z}$ are defined according to Eq.~(\ref{eq:landau}) with \mbox{$m_q = \overline{m}_q(\mu)$}:
\begin{align}
h &= \bar{h} - 2 \, \Delta \, \bar{h} \, \frac{\bar{h}-1}{\bar{h}+1} \,, \nonumber \\
z &= \bar{z} - 2 \, \Delta \, \bar{z} \, \frac{\bar{z}-1}{\bar{z}+1} \,.
\end{align}

The amplitudes in the schemes (a), (b) and (c) have a common polynomial structure, which contains only a limited number of combinations of the denominators~(\ref{eq:denom}):
\begin{align}
A^{(2)}_q = 16\pi^2\,S_\e^2\,m_q^3\,y_q\,v\,C_F\,\cdot &\left[ \frac{c_1}{l_1} + \frac{c_2}{l_1 \, l_5} + \frac{c_3}{l_1 \, l_6} + \frac{c_4}{l_1 \, l_6^2} + \frac{c_5}{l_1 \, l_9 \, l_{10}} + \frac{c_6}{l_2 \, l_4} + \frac{c_7}{l_3 \, l_4} + \frac{c_8}{l_4} \right. \nonumber \\
	&\left.\; + \frac{c_9}{l_4 \, l_8} + \frac{c_{10}}{l_4 \, l_{11}} + \frac{c_{11}}{l_5 \, l_7} + \frac{c_{12}}{l_5 \, l_8} + \frac{c_{13}}{l_7} + \frac{c_{14}}{l_9} + \frac{c_{15}}{l_{10}}  + \frac{c_{16}}{l_{12}} \nonumber \right] \nonumber \\
	&+ {\cal O}(\eps)\,.
\label{eq:NLO}
\end{align}
The coefficients $c_i$ are linear combinations of GHPLs multiplied by some power of $h$ or $z$ with positive exponent. The complete analytic expression of the two-loop amplitude exceeds the scope of this paper and is 
attached to the arXiv submission of this article.

The analytic result of the full two-loop amplitude enables us to derive its limit for small quark masses and gain information about its logarithmic structure. We perform this expansion by removing trailing zeros in the indices of the GHPLs using the shuffle relation so that the logarithmic singularities become explicit. Subsequently, we apply the scaling relation to the remaining GHPLs of the form~(\ref{eq:alphabet}), which turn into
\begin{align}
G\left(a_1,\ldots,a_n;1\right) \quad &\text{with} \quad a_i \in \{ 0,\pm\frac{1}{h},\frac{z}{h},\frac{1}{z\,h},\frac{J_z}{h},\frac{1}{J_z\,h},\frac{K_z^\pm}{h}, \frac{L_z^\pm}{h} \} \,, \nonumber \\
G\left(b_1,\ldots,b_n;1\right) \quad &\text{with} \quad b_i \in \{ 0,\pm\frac{1}{z},\frac{c}{z},\frac{\bar{c}}{z} \} \,,
\end{align}
where $a_n \neq 0$ and $b_n \neq 0$. This procedure shifts the dependence on the quark mass from the argument to the indices of the GHPLs and allows expanding the integrand of their integral representation without the need to take care of the integration boundaries. Consequently, we solve the definition of the Landau variables in Eq.~(\ref{eq:landau}) for $h$ and $z$, replace them in the integral representation and expand the result in $m_q$.\\
A subtlety occurs when GHPLs of the form~(\ref{eq:nontrans}) are analyzed for small quark masses. The particular case $w_1=L_z^-$ gives rise to further singularities due to
\begin{equation}
\lim_{m_q \to 0} L_z^{-} = z +  \mathcal{O}(z^2) \,,
\end{equation}
which is why the corresponding GHPLs are isolated with the help of the shuffle relation. Finally, we expand the remaining GHPLs with $w_i=\{K_z^\pm,L_z^\pm\} \, (i \neq 1)$ in small quark masses, apply transformations of the type~(\ref{eq:trafo1}) and treat the resulting GHPLs in the same way as the ones above. In doing so, we obtain the following expression for the amplitude from Eq.~(\ref{eq:amp}) in the limit of small quark masses, renormalized in the OS scheme:
\begin{align}
\lim_{M_q \to 0} A_q^{(a)}(m_H,m_Z,M_q) &= 4\,i\,S_\e\,M_q\,Y_q\,v\,\left(m_H^2-m_Z^2\right)\,{\rm log}\left(\frac{M_q^2}{m_Z^2}\right)\,{\rm log}\left(\frac{m_Z^2}{m_H^2}\right) \label{eq:exp} \\
&\quad \left[1 + C_F\,\frac{\alpha_s(\mu)}{\pi} \, {\rm log}\left(\frac{M_q^2}{m_Z^2}\right) \right. \nonumber \\
&\quad\left.\left\{ \frac{3}{4} - \frac{1}{8} \, {\rm log}\left(\frac{m_Z^2}{m_H^2}\right) + \frac{1}{2} \, {\rm log}\left(1-\frac{m_Z^2}{m_H^2}\right) + \frac{{\rm Li_2}\left(\frac{m_Z^2}{m_H^2}\right) - \zeta_2}{2 \, {\rm log}\left(\frac{m_Z^2}{m_H^2}\right)} \right\} \right] \nonumber \,.
\end{align}
To check the expansions, we validated numerically (using the \textsc{GiNaC}~\cite{Vollinga:2004sn,Bauer:2000cp} implementation) that the individual GHPLs converge towards their expansions in the limit of small quark masses. 
In addition, we rederived the corresponding limit of the $H\to\gamma\gamma$ amplitude by starting from the $H\to Z\gamma$ amplitude and setting the $Z$ boson mass to zero. With the manipulations described above, we agree with the previously available ratio of the two-to-one-loop amplitude for the process $H\to\gamma\gamma$ in small quark masses \cite{Spira:1995rr,Akhoury:2001mz}. 

It is important to note that Eq.~(\ref{eq:exp}) contains only single-logarithmic terms, which is in contrast to the $H\to\gamma\gamma$ case. The non-trivial cancellation of the double-logarithmic terms, which originate from the Sudakov region,
  takes place both in the one-loop and in the two-loop amplitudes of Eqs.~(\ref{eq:LO})~and~(\ref{eq:NLO}), leaving only single-logarithmic terms at each order. A double-logarithmic Sudakov resummation, as performed for $H\to\gamma\gamma$ in 
  \cite{Akhoury:2001mz} is therefore not needed for $H\to Z\gamma$.  We observe from Eq.~(\ref{eq:exp}) that the introduction of a running Yukawa coupling, scheme (b), resums the single-logarithmic contribution that is independent of $m_Z^2/m_H^2$. 
    
\section{Numerical results}
\label{sec:results}

The calculation of the master integrals and the amplitude outlined in Section~\ref{sec:calc} is performed for the values
\begin{equation}
0 < h < 1 \,, \quad 0 < z < 1 \,.
\end{equation}
This corresponds to the Euclidean region, where $m_H^2$ and $m_Z^2$ in Eq.~(\ref{eq:landau}) are negative and the master integrals are real. In order to get a physical expression, the results have to be analytically continued to the physical Minkowski region, where we distinguish three kinematic regions:
\begin{itemize}
\item \makebox[2cm][l]{Region I:} $m_Z^2 < m_H^2 < 4 \, m_q^2$ \,,
\item \makebox[2cm][l]{Region II:} $m_Z^2 < 4 \, m_q^2 < m_H^2$ \,,
\item \makebox[2cm][l]{Region III:} $4 \, m_q^2 < m_Z^2 < m_H^2$ \,.
\end{itemize}
In Region~I, the virtualities of the external massive particles are below the threshold induced by the particle running in the loop and the amplitude is real~\cite{'tHooft:1978xw}. The two solutions of each variable in Eq.~(\ref{eq:landau}) become imaginary in this region with
\begin{align}
h &= {\rm e}^{\pm2i\phi_H} \,, \\
z &= {\rm e}^{\pm2i\phi_Z} \,.
\end{align}
$\phi_H$ and $\phi_Z$ are phase factors given by
\begin{equation}
\phi_i = {\rm arctan} \sqrt{\frac{m_i^2}{4 m_q^2 - m_i^2}} \,,
\end{equation}
i.e. the variables lie on the unit circle in the complex plane and the second solution is the complex conjugated of the first one. In general, care has to be taken when choosing one of them such that it is in agreement with the common $+i0$ prescription for the Mandelstam variables $m_H^2$ and $m_Z^2$. Due to the real amplitude, however, there is no ambiguity in this case and the imaginary parts of $h$ and $z$ can be chosen freely.

In Region III, $m_H^2$ and $m_Z^2$ are beyond the threshold and the amplitude picks up an imaginary part. The two solutions of each variable in Eq.~(\ref{eq:landau}) are real, with one fulfilling
\begin{equation}
-1 < h < 0 \,, \quad -1 < z < 0 \,,
\end{equation}
and the other one being its inverse. In contrast to Region I, one has to be careful when assigning a small imaginary part to $h$ and $z$ in order to fix branch cut ambiguities related to the Mandelstam variables. In this case, a positive imaginary part leads to the correct result if $\left|h\right| < 1$ or $\left|z\right| < 1$ and a negative imaginary part has to be applied when $\left|h\right| > 1$ or $\left|z\right| > 1$.

From the input parameters specified below, it is obvious that the top quark amplitude is calculated in Region I, while the bottom quark amplitude is computed in Region III. Region II  is not needed for the physical values of the masses.

Since the analytical expression for $A^{(2)}_q$ is given in terms of GHPLs, it can be evaluated using \textsc{GiNaC}~\cite{Vollinga:2004sn,Bauer:2000cp}. For masses and couplings, we use the input values
\begin{alignat}{4}
\alpha_s^{(5)} (m_H) &= 0.1130114 \, , &\quad \alpha &= 1/128 \, , &\quad G_F &= 1.1663787 \cdot 10^{-5} \, \mathrm{GeV}^{-2} \, , \notag \\
m_H &= 125.7 \, \mathrm{GeV} \, , &\quad m_Z &= 91.1876 \, \mathrm{GeV} \, , &\quad m_W &= 80.385 \, \mathrm{GeV} \, , \notag \\
M_t &= 173.21 \, \mathrm{GeV} \, , &\quad \overline{m}_t(m_H) &= 167.21 \, \mathrm{GeV} \, , &\quad M_b &= 4.7652 \, \mathrm{GeV} \, , \notag \\
\overline{m}_b(m_H) &= 2.7832 \, \mathrm{GeV} \, , &\quad \mathrm{sin}^2 \, \theta_w &= 0.23126 \, , &\quad \mathrm{cos}^2 \, \theta_w &= 0.76874 \, , \notag \\
Q_t &= 2/3 \, , &\quad Q_b &= -1/3 \, , &\quad I_t^3 &= 1/2 \, , \notag \\
I_b^3 &= -1/2 \, , &\quad C_F &= 4/3 \, , &\quad N_c &= 3 \, .
\end{alignat}
They were obtained by evolving the values of the Particle Data Group Collaboration~\cite{Agashe:2014kda} to the scale $\mu=m_H$ with the two-loop renormalization group equations~\cite{Chetyrkin:2000yt}.
To resum potentially large single logarithms in $\overline{m}_q/m_H$ to all orders in the perturbative expansion, we use the two-loop renormalization group equations~\cite{Chetyrkin:2000yt} to evolve the $\overline{{\rm MS}}$ quark mass (and accordingly the Yukawa coupling) from the matching scale to $\mu=m_H$. This leads to the following next-to-leading-order decay width $\Gamma^{(2)}$ in the renormalization schemes (a), (b) and (c):
\begin{align}
\Gamma^{(2,a)} &\overset{\textcolor{white}{\mu=m_H}}{=} \left[ 7.07533 + 0.42800 \, \frac{\alpha_s(\mu)}{\pi} \right] {\rm keV} \nonumber \\
	&\overset{\mu=m_H}{=} 7.09072 \, {\rm keV} \,, \\
\Gamma^{(2,b)} &\overset{\mu=m_H}{=} \left[ 7.09409 + \frac{\alpha_s(m_H)}{\pi} \left( -0.53266 - 0.76661 \, \log \frac{m_H^2}{\overline{m}_t^2(m_H)} + 0.01229 \, \log \frac{m_H^2}{\overline{m}_b^2(m_H)} \right) \right] {\rm keV} \nonumber \\
	&\overset{\textcolor{white}{\mu=m_H}}{=} 7.09403 \, {\rm keV} \,, \label{eq:widthhyb} \\
\Gamma^{(2,c)} &\overset{\mu=m_H}{=} \left[ 7.05934 + \frac{\alpha_s(m_H)}{\pi} \left( 0.64587 + 0.10597 \, \log \frac{m_H^2}{\overline{m}_t^2(m_H)} + 0.01453 \, \log \frac{m_H^2}{\overline{m}_b^2(m_H)} \right) \right] {\rm keV} \nonumber \\
	&\overset{\textcolor{white}{\mu=m_H}}{=} 7.08438 \, {\rm keV} \,. \label{eq:widthms}
\end{align}
The breakup of the terms in the first lines of Eqs.~(\ref{eq:widthhyb}) and (\ref{eq:widthms}) is to illustrate the relative numerical importance of the individual contributions.

\begin{figure}[t]
\begin{center}
\includegraphics[width=0.92\textwidth]{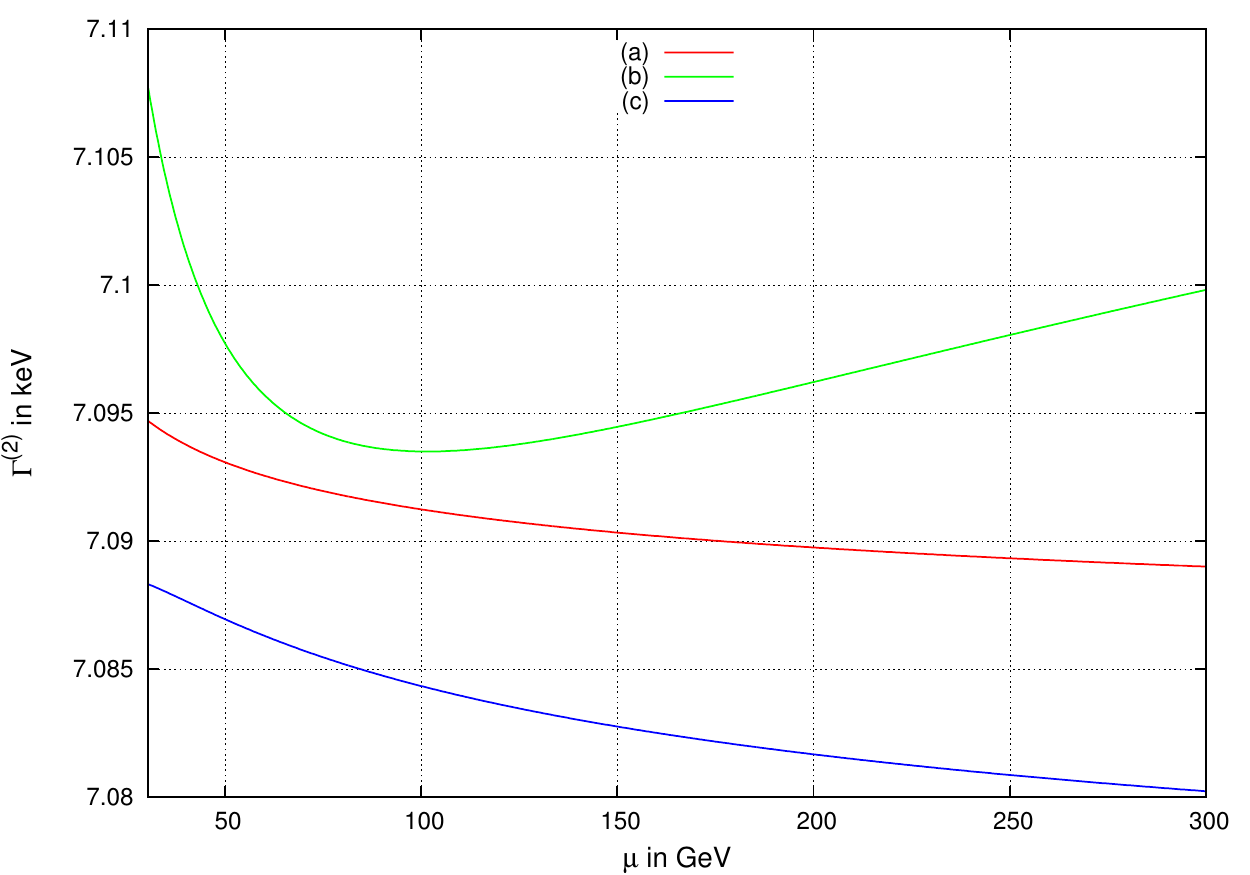}
\caption{Scale variation of the next-to-leading-order decay width $\Gamma^{(2)}$ in the renormalization schemes (a), (b) and (c) for $30\,{\rm GeV}<\mu<300\,{\rm GeV}$.}
\label{fig:varscale}
\end{center}
\end{figure}

An estimation of the uncertainty on the prediction from missing higher orders is provided by varying $m_H/2<\mu<2\,m_H$ in Fig.~\ref{fig:varscale}. For every data point $\mu=\mu_0$, the $\overline{{\rm MS}}$~quark mass $\overline{m}_q(\mu_0)$ and the strong coupling constant $\alpha_s(\mu_0)$ are evolved to the scale $\mu_0$ using the two-loop renormalization group equations~\cite{Chetyrkin:2000yt}.
The leading-order decay width $\Gamma^{(1)}$ and the next-to-leading-order decay width $\Gamma^{(2)}$ can be separated into contributions from the $W$~boson, top quark and bottom quark amplitudes as well as their interferences. The corresponding values are shown in Table~\ref{tab:width}, from which it becomes clear that the bottom quark amplitude has to be taken into account, since its interference with the $W$ amplitude is of the same order of magnitude as the self-interference of the top quark amplitude at one loop. Moreover, the combination $\Gamma^{(2)}_{Wb}$ exceeds the top quark self-interference $\Gamma^{(2)}_{tt}$ at two loops.
\begin{table}[b]
\caption{Various contributions to the numerical result of the leading-order decay width $\Gamma^{(1)}$ and the next-to-leading-order decay width $\Gamma^{(2)}$ in the renormalization schemes (a), (b) and (c), evaluated for $\mu=m_H$. In case of $\Gamma^{(1)}_{ij}$, the subscripts $i$ and $j$ indicate the interference of the one-loop amplitudes $A^{(1)}_W$, $A^{(1)}_t$ and $A^{(1)}_b$, whereas $\Gamma^{(2)}_{ij}$ describes the interference of the one-loop amplitude $A^{(1)}_i$ with the two-loop amplitude $A^{(2)}_j$. All values are given in keV.}
\setlength{\tabcolsep}{0.9cm}
\begin{tabularx}{\textwidth}{lrrr}
\toprule
Partial width & \multicolumn{1}{c}{(a)} & \multicolumn{1}{c}{(b)} & \multicolumn{1}{c}{(c)} \\
\midrule
$\Gamma^{(1)}_{WW}$ & $7.86845996$ & $7.86845996$ & $7.86845996$ \\
$\Gamma^{(1)}_{Wt}$ & $-0.83636436$ & $-0.80736905$ & $-0.84015333$ \\
$\Gamma^{(1)}_{Wb}$ & $0.02216139$ & $0.01294390$ & $0.00908488$ \\
$\Gamma^{(1)}_{tt}$ & $0.02222498$ & $0.02071068$ & $0.02242680$ \\
$\Gamma^{(1)}_{tb}$ & $-0.001177803$ & $-0.00066408$ & $-0.00048502$ \\
$\Gamma^{(1)}_{bb}$ & $0.00002103$ & $0.00000717$ & $0.00000325$ \\
\midrule
$\Gamma^{(1)}$ & $7.07532519$ & $7.09408860$ & $7.05933655$ \\
\midrule
$\Gamma^{(2)}_{Wt}$ & $0.02213199$ & $-0.00078617$ & $0.02467587$ \\
$\Gamma^{(2)}_{Wb}$ & $-0.00588750$ & $0.00073044$ & $0.00176120$ \\
$\Gamma^{(2)}_{tt}$ & $-0.00117624$ & $0.00004033$ & $-0.00131738$ \\
$\Gamma^{(2)}_{tb}$ & $0.00031290$ & $-0.00003747$ & $-0.00009403$ \\
$\Gamma^{(2)}_{bt}$ & $0.00003117$ & $-0.00000065$ & $0.00001425$ \\
$\Gamma^{(2)}_{bb}$ & $-0.00001592$ & $-0.00000081$ & $0.00000078$ \\
\midrule
$\Gamma^{(2)}$ & $7.09072159$ & $7.09403427$ & $7.08437723$ \\
\bottomrule
\label{tab:width}
\end{tabularx}
\end{table}

Our on-shell results are in agreement with the numerical findings of~\cite{Spira:1991tj}. Furthermore, we performed a detailed numerical comparison with~\cite{Bonciani:2015} and found agreement to high precision.

We observe that the NLO results for the decay width are consistent between the three schemes. The relative size of the 
NLO correction is 2$\permille$ in scheme (a), below $10^{-5}$ in scheme (b) and 3$\permille$ in scheme (c). The 
very small corrections in scheme (b) are however in large part due to numerical 
cancellations between a priori unrelated terms.
The spread between
the different schemes is 1.3$\permille$ at $\mu=m_H$, and variations of the renormalization scale change the 
predictions in either given scheme by at most 0.4$\permille$.

\section{Conclusions}
\label{sec:conc}
In this paper, we have revisited the QCD corrections to the rare loop-induced Higgs boson decay $H\to Z\gamma$.
The relevant two-loop three-point integrals with two different external masses and one internal mass were derived 
analytically, using a reduction to master integrals, which were then computed using differential equations. These 
integrals are also an important ingredient to the two-loop amplitudes for Higgs-plus-jet production in gluon fusion with 
full dependence on the internal quark masses. 

By expanding the one-loop and two-loop matrix elements in the on-shell scheme
 for  $H\to Z\gamma$ in the limit of small quark masses, we noted 
the absence of double-logarithmic contributions, which is in contrast to $H\to \gamma\gamma$~\cite{Spira:1995rr,Akhoury:2001mz}; single-logarithmic terms are resummed by the introduction of a running Yukawa coupling.

We investigated the  dependence of the corrections on the 
renormalization scheme used for the quark mass and Yukawa coupling. We observe that 
the results for the decay rate  in on-shell and $\overline{{\rm MS}}$ schemes, 
as well as in a hybrid scheme with on-shell mass and $\overline{{\rm MS}}$
Yukawa coupling, are well consistent with each other, and that corrections are in the sub-per-cent range in all three 
schemes (being smallest in the hybrid scheme). We confirm the previously available
numerical on-shell result~\cite{Spira:1991tj} and agree with an independent caclulation~\cite{Bonciani:2015}.
The residual QCD  uncertainty on the   $H\to Z\gamma$ decay rate
is around 1.7$\permille$ from the combination of scale variation
and  spread between the different renormalization schemes.

\section*{Acknowledgements}
We are grateful to R. Bonciani for providing results to compare with their independent calculation prior to publication~\cite{Bonciani:2015}. In addition, we thank M.~Wiesemann for useful comments on the renormalization, L.~Tancredi for enlightening conversations on many different issues of this project and A.~von~Manteuffel for useful comments on the manuscript. Finally, we wish to thank S.~Borowka, A.~von~Manteuffel and E.~Weihs for their assistance with the use of \textsc{SecDec}, \textsc{Reduze2} and the inhouse \textsc{Mathematica} package, respectively. This research was supported in part by the Swiss National Science Foundation (SNF) under contract 200020-149517, as well as  by the European Commission through the  ERC Advanced Grant ``MC@NNLO" (340983). The Feynman graphs in this paper have been drawn with \textsc{Jaxodraw}~\cite{Vermaseren:1994je, Binosi:2008ig}.

\begin{appendix}
\allowdisplaybreaks
\section{Canonical Master integrals}
\label{sec:appendix}

In this appendix, we provide the relations between the two-loop canonical master integrals that appear in Eq.~(\ref{eq:canon}) and the Laporta master integrals of Fig.~\ref{fig:master} in terms of the Landau variables defined in Eq.~(\ref{eq:landau}). The canonical integrals are normalized such that their Laurent expansion starts at order $\e^0$. In addition, we extract the mass dimension of the integrals and obtain
\begin{align}
M_1 &= \e^2 \, I_1 \,, \nonumber \\
M_2 &= \e^2 \, m_q^2 \, \frac{(z+1) (z-1)}{z} \, I_2 \,, \nonumber \\
M_3 &= \e^2 \, m_q^2 \, \frac{z-1}{z} \, \left[(z+1) \, I_3 + I_4 \right] \,, \nonumber \\
M_4 &= \e^2 \, m_q^2 \, \frac{(z-1)^2}{z} \, I_4 \,, \nonumber \\
M_5 &= \e^2 \, m_q^2 \, \frac{(h+1) (h-1)}{h} \, I_5 \,, \nonumber \\
M_6 &= \e^2 \, m_q^2 \, \frac{h-1}{h} \, \left[(h+1) \, I_6 + I_7 \right] \,, \nonumber \\
M_7 &= \e^2 \, m_q^2 \, \frac{(h-1)^2}{h} \, I_7 \,, \nonumber \\
M_8 &= \e^2 \, m_q^4 \, \frac{(z+1)^2 (z-1)^2}{z^2} \, I_8 \,, \nonumber \\
M_9 &= - \e^3 \, m_q^2 \, \frac{(h-z) (h z-1)}{h z} \, I_9 \,, \nonumber \\
M_{10} &= - \e^3 \, m_q^2 \, \frac{(h-z) (h z-1)}{h z} \, I_{10} \,, \nonumber \\
M_{11} &= \e^2 \, m_q^4 \, \frac{(h^2-1) (z^2-1)}{h z} \, I_{11} \,, \nonumber \\
M_{12} &= \e^2 \, m_q^4 \, \frac{(h+1)^2 (h-1)^2}{h^2} \, I_{12} \,, \nonumber \\
M_{13} &= - \e^3 \, m_q^2 \, \frac{(h-z) (h z-1)}{h z} \, I_{13} \,, \nonumber \\
M_{14} &= - \e^3 \, m_q^2 \, \frac{(h-z) (h z-1)}{h z} \, I_{14} \,, \nonumber \\
M_{15} &= - \e^2 \, m_q^2 \, \frac{z\,(h^2+1) - h\,(z+1)}{2\,z\,(h^2+1) - h\,(z+1)^2} \, \left[ -\frac{3}{2} \, \frac{(h-1)^2}{h} I_{7} + \eps \, \frac{(h-z) (h z-1)}{h z} \left(2\,I_{13} + I_{14}\right) \right. \nonumber \\
	&\left.\qquad\qquad\qquad\qquad\qquad\qquad\qquad\quad\;\;\;+ m_q^2 \, \frac{(z^2-1) (h^2+1-h\,(z+1))}{h z} \, I_{15} \right] \,, \nonumber \\
M_{16} &= - \e^3 \, m_q^2 \, \frac{(h-z) (h z-1)}{h z} \, I_{16} \,, \nonumber \\
M_{17} &= - \e^3 \, m_q^2 \, \frac{(h-z) (h z-1)}{h z} \, I_{17} \,, \nonumber \\
M_{18} &= - \e^2 \, m_q^2 \, \frac{h\,(z^2+1) - z\,(h+1)}{2\,h\,(z^2+1) - z\,(h+1)^2} \, \left[ -\frac{3}{2} \, \frac{(z-1)^2}{z} I_{4} - \eps \, \frac{(h-z) (h z-1)}{h z} \left(2\,I_{16} + I_{17}\right) \right. \nonumber \\
	&\left.\qquad\qquad\qquad\qquad\qquad\qquad\qquad\quad\;\;\;+ m_q^2 \, \frac{(h^2-1) (z^2+1-z\,(h+1))}{h z} \, I_{18} \right] \,, \nonumber \\
M_{19} &= \e^3 \, m_q^4 \, \frac{(h-z) (h z-1) (z^2-1)}{h z^2} \, I_{19} \,, \nonumber \\
M_{20} &= - \e^3 \, m_q^4 \, \frac{(h-z) (h z-1) (h^2-1)}{h^2 z} \, I_{20} \,, \nonumber \\
M_{21} &= - \e^4 \, m_q^2 \, \frac{(h-z) (h z-1)}{h z} \, I_{21} \,, \nonumber \\
M_{22} &= - \e^3 \, m_q^4 \, \frac{(h-z) (h z-1) (h^2-1)}{h^2 z} \, I_{22} \,, \nonumber \\
M_{23} &= - \e^4 \, m_q^2 \, \frac{(h-z) (h z-1)}{h z} \, I_{23} \,, \nonumber \\
M_{24} &= - \e^3 \, m_q^4 \, \frac{(h-z) (h z-1) (z^2-1)}{h z^2} \, I_{24} \,, \nonumber \\
M_{25} &= \e^4 \, m_q^2 \, \frac{(h-z) (h z-1)}{h z} \, I_{25} \,, \nonumber \\
M_{26} &= \e^3 \, m_q^4 \, \frac{(h-z) (h z-1) (z^2-1)}{h z^2} \, I_{26} \,, \nonumber \\
M_{27} &= \e^3 \, m_q^4 \, \frac{(h-z) (h z-1) (h^2-1)}{h^2 z} \, I_{27} \,, \nonumber \\
M_{28} &= \e^2 \, m_q^4 \, \frac{h z-1}{h z} \, \left[ -4 \, (h z-1) \, I_{11} + 2 \, \e \, (h-z) \, \left(\frac{z-1}{z} \, I_{26} - \frac{h-1}{h} \, I_{27} \right) \right. \nonumber \\
	&\qquad\qquad\qquad\quad\;\;\left.+ m_q^2 \, \frac{(h-z)^2 (h z-1)}{h z} \, I_{28} \right] \,.
\label{eq:appendix}
\end{align}
$I_1$ is the two-loop tadpole integral with both propagators taken to be squared. It is given by
\begin{equation}
I_1 = \int \frac{{\rm d}^Dk}{(2\pi)^D} \int \frac{{\rm d}^Dl}{(2\pi)^D} \frac{1}{(k^2-m_q^2)^2 \, (l^2-m_q^2)^2} = \frac{S_\e^2}{\e^2}
\end{equation}
so that
\begin{equation}
M_1 = S_\e^2 \,,
\end{equation}
where $S_\e^2$ is the common normalization factor of all two-loop master integrals defined in Eq.~(\ref{eq:norm}).

\end{appendix}

\end{document}